\documentclass[conference,letterpaper]{IEEEtran}
\IEEEoverridecommandlockouts
\usepackage{cite}
\usepackage{amsmath,amssymb,amsfonts}
\usepackage{algorithmic}
\usepackage{graphicx}
\usepackage{textcomp}
\usepackage{xcolor}
\usepackage{tcolorbox}
\usepackage{subcaption}
\usepackage{quotes}
\usepackage{hyperref}
\usepackage{array}
\usepackage{multirow}
\usepackage{booktabs}
\usepackage{float}
\usepackage{listings}
\usepackage{xurl}
\usepackage{soul}

\def\cameraready{1}

\newcommand{\orange}[1]{\ifnum\cameraready=1 #1\else\textcolor{orange}{#1}\fi}
\def\BibTeX{{\rm B\kern-.05em{\sc i\kern-.025em b}\kern-.08em
    T\kern-.1667em\lower.7ex\hbox{E}\kern-.125emX}}
\begin{document}

\title{Toward Reproducible and Standardized Computer Architecture Simulation with gem5\\
}


\author{
Kunal Pai\textsuperscript{*}, Harshil Patel\textsuperscript{*}, Erin Le, Noah Krim\textsuperscript{\dag}, Mahyar Samani, Bobby R. Bruce, and Jason Lowe-Power\\
\textit{University of California, Davis}, Davis, USA \\
\{kunpai, hpppatel, ejle, msamani, bbruce, jlowepower\}@ucdavis.edu, nkrim62@gmail.com
\thanks{\textsuperscript{*}These authors contributed equally to this work.}
\thanks{\textsuperscript{\dag}Work completed while at the University of California, Davis.}
}



\maketitle
\thispagestyle{empty}
\pagestyle{empty}
\begin{abstract}
Reproducibility in simulation-based computer architecture research requires coordinating artifacts like disk images, kernels, and benchmarks, but existing workflows are inconsistent.
We improve gem5, an open-source simulator with over 1600 forks, and gem5 Resources, a centralized repository of over 2000 pre-packaged artifacts, to address these issues.
While gem5 Resources enables artifact sharing, researchers still face challenges.
Creating custom disk images is complex and time-consuming, with no standardized process across ISAs, making it difficult to extend and share images.
gem5 provides limited guest-host communication features through a set of predefined exit events that restrict researchers' ability to dynamically control and monitor simulations.
Lastly, running simulations with multiple workloads requires researchers to write custom external scripts to coordinate multiple gem5 simulations which creates error-prone and hard-to-reproduce workflows.
To overcome this, we introduce several features in gem5 and gem5 Resources.
We standardize disk-image creation across x86, ARM, and RISC-V using Packer, and provide validated base images with pre-annotated benchmark suites (NPB, GAPBS).
We provide 12 new disk images, 6 new kernels, and over 200 workloads across three ISAs.
We refactor the exit event system to a class-based model and introduce hypercalls for enhanced guest-host communication that allows researchers to define custom behavior for their exit events.
We also provide a utility to remotely monitor simulations and the gem5-bridge driver for user-space m5 operations.
Additionally, we implemented Suites and MultiSim to enable parallel full-system simulations from gem5 configuration scripts, eliminating the need for external scripting.
These features reduce setup complexity and provide extensible, validated resources that improve reproducibility and standardization.
\end{abstract}


\section{Introduction}\label{sec:introduction}

Reproducibility and usability remain major challenges in computer architecture research~\cite{antunes2024reproducibility}, particularly for simulation-based studies.
These challenges are often framed in terms of the ACM's definitions of reproducibility levels: repeatability (same team, same setup), replicability (different team, same setup), and reproducibility (different team, different setup)~\cite{acm_artifact_review_badging,plesser2018reproducibility}.
Achieving any level is difficult due to complex experimental workflows involving diverse toolchains, dependencies, and library configurations, which create substantial barriers to initial adoption and replication.
Insufficient documentation of experimental workflows and execution environments further compromises transparency and reproducibility.
The absence of standardized practices for sharing experimental artifacts affects verification and future research.
A lack of version control also makes it difficult to trace the exact state of a workflow.
These combined factors highlight the need for simulation infrastructure that is both robust and accessible, supported by well-documented, standardized methodologies to enable broader adoption and reliable reproduction of results.


We address these challenges within gem5, a widely-used open-source computer architecture simulator~\cite{binkert2011gem5,lowe2020gem5}.
gem5 is an event-driven, modular tool capable of simulating a diverse range of computer systems, from simple in-order processors to advanced out-of-order processors, as well as basic memory systems to complex memory hierarchies.
gem5 supports both full-system simulation and syscall emulation (SE) mode, providing flexibility for various use cases.
Beyond traditional CPU simulations, gem5 can model GPUs, network-on-chip (NoC) systems and emerging technology, broadening its applicability to different computing paradigms~\cite{pai2024potential,pai2025superconducting,fariborz2025nova}.
Written in C++ and Python, gem5 is highly configurable and extensible, supporting multiple ISAs, including x86, ARM, and RISC-V.
Researchers use gem5 in fields such as computer architecture, operating systems, compilers, and security.
In industry, gem5 enables rapid iteration in performance modeling, software development, and hardware design.

While gem5 offers a flexible simulation framework, setting up experiments using standard benchmarks and workloads can be complex, time-consuming and ad-hoc.
In response, gem5ART and gem5 Resources were developed to enable reproducibility in gem5~\cite{bruce2021enabling}.
gem5ART is a set of Python libraries that can be used to run gem5 experiments in a more structured way and allows storing results in a database for future use~\cite{bruce2021enabling}.
gem5 Resources is a centralized repository of artifacts that are not needed for gem5 to build but are used by researchers to run experiments. 
Since the release, gem5 Resources have been maintained and expanded.
These resources include ready-to-use benchmark suites (e.g., PARSEC~\cite{bienia2008parsec}, NAS Parallel Benchmarks (NPB)~\cite{npb}, GAP Benchmark Suite~\cite{beamer2015gap}, VRG microbenchmarks~\cite{nowatzki2015architectural}), workloads, and sampling techniques (e.g., checkpoints, SimPoints~\cite{sherwood2002automatically,kunal2023matchedposter}, LoopPoints~\cite{sabu2022looppoint,Qiu2023LoopPoint}) provided in formats consumable by gem5.

While gem5ART and gem5 Resources have simplified artifact discovery and management, key usability and reproducibility gaps remain.
In this paper, we address three such challenges through contributions that have been merged upstream in the gem5 v25.0 release:

\begin{itemize}

    \item \textbf{Workload Standardization:} Inconsistent disk image configurations and creation processes can lead to variable results.
    We standardize disk image creation by adopting a unified Packer-based workflow for x86, ARM, and RISC-V that uses the same modern Ubuntu LTS release and identical configuration parameters, abstracting hundreds of low-level, ISA-specific construction steps into declarative scripts.
    This standardization enables us to provide 12 new disk images, 6 new kernels, and over 200 widely-used benchmarks across three ISAs.
    We also validate these resources so that researchers can directly plug them into their experiments.

    \item \textbf{Decoupled Guest-Host Communication:} We refactored gem5's exit event system from a generator-based implementation to a class-based one and introduced hypercalls to decouple exit handling from workloads.
    This change allows researchers to create reusable and flexible exit handlers.

    \item \textbf{Parallel Experiment Orchestration:} Running many simulations across multiple configurations and benchmarks, in parallel, is fundamental to architecture research, yet coordinating these experiments has historically required separate initializations and external scripts that are difficult to share, maintain, and reproduce. 
    To address these challenges, we introduce a suite abstraction, which provides a structured way to define sets of workloads.
    We also add initial support for multi-simulation to streamline running multiple experiments directly from gem5 configuration scripts. 
    Together, these features aim to simplify experiment setup and reduce reliance on external orchestration, helping make workflows more self-contained and reproducible.


\end{itemize}

Importantly, our contributions do not alter gem5's core simulation logic.
The sole architectural addition is a lightweight pseudo-instruction for host communication.
Therefore, we preserve simulated performance while improving usability and experiment reproducibility.

\section{Background and Motivation}\label{sec:motivation}

gem5 is a widely-used, cycle-level, open-source simulator for computer architecture research~\cite{binkert2011gem5, lowe2020gem5}.
With more than 1,600 GitHub forks and contributions from both academia and industry, it supports a broad range of systems, from simple in-order cores to complex out-of-order processors and advanced memory hierarchies.

Because architectural research typically requires evaluating designs across many workloads, gem5 enables simulation of diverse applications and benchmark suites in either syscall emulation (SE) mode or full-system (FS) mode.
These workloads range from small microbenchmarks to being part of larger suites such as PARSEC~\cite{bienia2008parsec}, NAS Parallel Benchmarks (NPB)~\cite{npb}, and GAPBS~\cite{beamer2015gap}.
In gem5, each such unit is represented as a ``Workload,'' which packages all required dependencies, including disk images, kernels, and input parameters.

A ``Workload'' encapsulates only a single benchmark or application.
This reflects gem5's design: one simulation per gem5 instance.
Therefore, researchers who needed to run a suite of benchmarks had to write custom external scripts, typically Bash, to orchestrate multiple gem5 simulations.
They also had to manage separate output directories to avoid overwriting the default \texttt{m5out}.
This workflow was cumbersome and error-prone, and created unnecessary friction for both new and experienced users.

The notion of a self-contained ``Workload'' originally applied only to artifacts officially curated in gem5 Resources~\cite{gem5_resources}.
Users who wanted to run custom applications or benchmarks had to manually ``bundle'' their workloads inside gem5 configuration scripts.
Before our changes, the only way to load non-standard artifacts was to instantiate a \texttt{CustomResource} and manually enumerate every component of the workload (binaries, disk images, kernels, metadata, and so on).
This process was verbose, error-prone and inconsistent across different research groups, since reproduction required reverse-engineering someone else's setup.

The reproducibility challenge was further compounded by different ISAs (x86, ARM, RISC-V) having different build systems.
Users also had to manually track which binaries, disk images, kernel versions, and input parameters were used for each experiment, often recording this information outside of gem5 in ad hoc notes or scripts.
Moreover, subtle differences in disk image internals, such as whether the system uses \texttt{systemd} during boot, whether the user has root privileges, or how initialization is configured, can influence simulation behavior and performance.
For example, we observed that the number of instructions in the region of interest (ROI) varied by 0.2\% to 1.3\% across NPB size S benchmarks on x86 when run with and without \texttt{systemd} during boot.
These inconsistencies make it difficult to reproduce results across systems, research groups, or even across runs on the same machine.



gem5 is also a discrete event simulator, where the simulated system (guest) and the host system run independently.
Historically, gem5 provided a generator-based exit event system to allow the guest to control the simulation by triggering predefined events such as checkpointing, regions of interest, or exiting the simulation.
Researchers could define custom handlers for each exit event type by writing generator functions that yield control back to the simulation after processing the event.

However, this exit event system tightly coupled the exit handlers and the workload.
Handlers had to process exit events in the exact order they were triggered by the workload.
Additionally, any change in the workload required modifying the handler to match the new order.
This tight coupling made handlers fragile and difficult to reuse across different workloads.


Secondly, exit events served as generic signals without conveying any simulation context.
For example, a common use case in gem5 is wanting to perform different actions when the same exit event occurs at different points in the simulation, such as printing a message and dumping statistics \textit{only} after the kernel finishes booting.
However, a workload might trigger multiple exit events of the same type at different stages in the simulation.
By design, the host could not tell why an exit event occurred or which point in the simulation it represented.
Since the host could not distinguish between them, researchers had to rely on fragile, order-dependent logic, and could not write fine-grained handlers that behaved differently depending on the simulation state.


Taken together, these limitations point to a clear need for improved \textit{workload standardization}, \textit{decoupled guest-host communication}, and robust \textit{parallel experiment orchestration} for gem5.

\section{Workload Standardization}\label{sec:disk-image-standardization}
Previously, there was no standardized process for creating disk images for gem5 full-system simulation.
For example, RISC-V and ARM disk images were built exclusively using QEMU, while x86 disk images required a combination of QEMU and Packer, each employing different methods to install Ubuntu.
These inconsistencies meant that researchers using different ISAs had to follow different procedures to create disk images, leading to potential variability in configurations such as installed packages, kernel versions, and exit events, which could affect simulation results.
To address these issues, we standardized disk image creation, kernels, pre-installed packages, and guest-host communication (Section~\ref{sec:guest-host-communication}).

We revamped the disk image creation process to address these inefficiencies and standardize the disk image creation workflow using Packer with its QEMU plugin.
Packer is an open-source infrastructure automation tool that enables the creation of consistent, reproducible machine images across multiple platforms from a single declarative specification, abstracting hundreds of low-level, platform-specific provisioning steps into a unified, version-controlled build process.
We leverage this same approach to provide ISA-agnostic scripts that produce gem5-ready images for x86, ARM, and RISC-V without ISA-specific procedures, eliminating the need for excessive manual adjustments.
We further improved this workflow by automatically installing the \texttt{gem5-bridge} kernel driver (Section~\ref{sec:gem5-bridge}) during disk image creation, ensuring all generated images support user-mode \texttt{m5} operations by default and are fully standardized for gem5.
Thus, all images share the same OS, kernel versions, runtime environments, pre-installed packages, and exit event configurations.
In addition, we standardize custom workloads, described in Section~\ref{sec:custom-workload-support}.

\subsection{Custom Workload Standardization}\label{sec:custom-workload-support}
We extended the \texttt{obtain\_resource} function to allow users to load their local resources directly~\cite{gem5_local_resources_support_real}, effectively deprecating the need for \texttt{CustomResource}, as discussed in Section~\ref{sec:motivation}.
This brings local resource loading in line with the standard mechanism used for gem5-hosted resources, enabling both to be accessed through the same unified API.
As a result, gem5-managed resources and user-provided resources can now be indexed, queried, and consumed uniformly.

A caveat is that local resources must follow the same schema as resources in the gem5 Resources database, and any dependency relationships (e.g., the components that constitute a Workload) must be specified manually.
A sample JSON definition for a local binary resource is shown below:

\begin{lstlisting}[language=Python,
    basicstyle=\ttfamily\footnotesize,
    breaklines=true,
    breakatwhitespace=true
]
[
    {
        "category": "binary",
        "id": "test-binary",
        "description": "Test",
        "architecture": "<ISA>",
        "url": "file:///path/to/binary",
        "resource_version": "1.0.0",
        "gem5_versions": [
            "25.0"
        ]
    }
]
\end{lstlisting}

\subsection{Runtime Standardization}\label{sec:gem5-bridge}
We introduced a new Linux kernel module, \texttt{gem5-bridge} to facilitate user-space execution of \texttt{m5} operations without requiring superuser privileges.\footnote{\url{https://github.com/gem5/gem5/pull/1480}}
~\texttt{m5} operations are pseudo-instructions that enable guest (simulated system) to host (simulator) communication during gem5 simulations.
Previously, invoking \texttt{m5} operations required \texttt{/dev/mem} access, which necessitates root privileges in the guest, forcing all benchmarks to run as root inside the simulation.
Invoking gem5 operations via privileged paths (e.g., \texttt{/dev/mem}) versus an unprivileged MMIO interface exercises different kernel and virtualization paths, and can mask security bugs, potentially affecting experimental validity and simulated stats.
The \texttt{gem5-bridge} driver eliminates this requirement by providing unprivileged guest access to \texttt{m5} operations.

Specifically, the module introduces a character device, \texttt{/dev/gem5\_bridge}, to provide controlled access to the \texttt{m5} MMIO range, while strictly validating requests to prevent the exposure of arbitrary memory.
By replacing unrestricted \texttt{/dev/mem} accesses with this targeted device, users can execute \texttt{m5} operations and run annotated benchmarks entirely in user mode.
Therefore, the driver safely enables \texttt{m5} operations in non-privileged environments and standardizes the runtime interface.

\begin{figure}[htbp]
\begin{lstlisting}[
    basicstyle=\ttfamily\footnotesize,
    breaklines=true,
    breakatwhitespace=true
]
/home/gem5/NPB3.4-OMP/bin/bt.S.x
 --------------------- M5 INIT --------------------- 
Can't open /dev/mem: Permission denied
\end{lstlisting}
\caption*{(a) Output of gem5 without the \texttt{gem5-bridge}
driver. The command would need to access \texttt{/dev/mem},
requiring root privileges.}

\begin{lstlisting}[
    basicstyle=\ttfamily\footnotesize,
    breaklines=true,
    breakatwhitespace=true
]
/home/gem5/NPB3.4-OMP/bin/bt.S.x
 --------------------- M5 INIT --------------------- 


 NAS Parallel Benchmarks (NPB3.4-OMP) - BT Benchmark

 No input file inputbt.data. Using compiled defaults
 Size:   12x  12x  12
 Iterations:   60       dt:   0.0100000
 Number of available threads:     2

 -------------------- ROI BEGIN -------------------- 
 Time step    1
 Time step   20
 Time step   40
 Time step   60
 -------------------- ROI END -------------------- 
\end{lstlisting}
\caption*{(b) Output of gem5 with the \texttt{gem5-bridge} driver.}

\caption{Comparison of gem5 output with and without the
\texttt{gem5-bridge} driver installed in the disk image. With
the driver, user-mode m5 operations work without root
privileges.}
\label{fig:gem5-bridge}
\end{figure}

  
  


\begin{figure*}
    \centering
    \begin{subfigure}{0.9\textwidth}
        \centering
        \includegraphics[width=\textwidth]{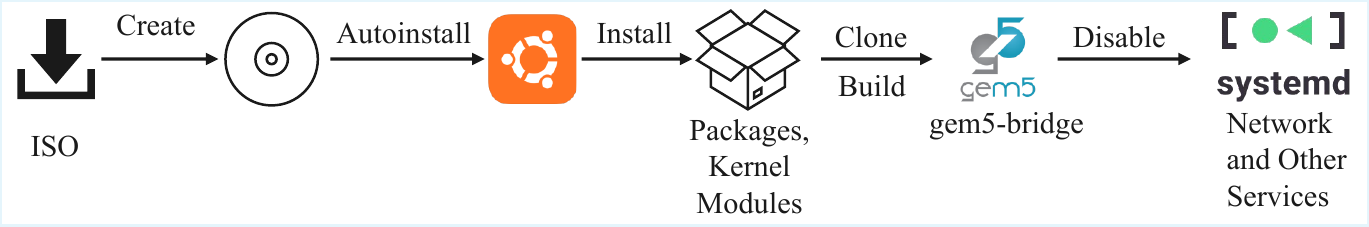}
        \caption{Disk Image Creation Workflow}
        \label{fig:disk-image-creation-workflow}
    \end{subfigure}
    \hfill
    \begin{subfigure}{0.9\textwidth}
        \centering
        \includegraphics[width=\textwidth]{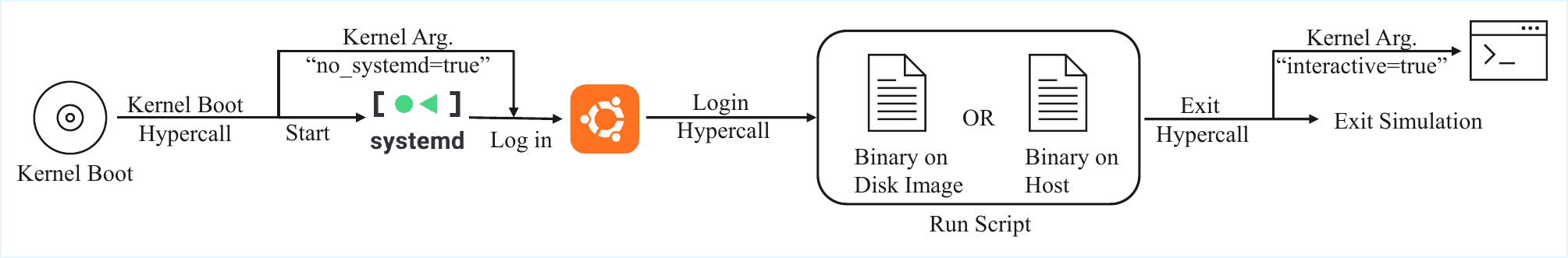}
        \caption{Hypercalls in Boot Workflow}
        \label{fig:hypercalls-boot-workflow}
    \end{subfigure}
    \caption{Disk Image Creation and Hypercalls Workflow}
    \label{fig:disk-image-hypercalls}
\end{figure*}

\subsection{Disk Image Workflow Standardization}\label{sec:disk-image-workflow-standardization}
The new disk images in gem5 are now created using an updated workflow that fully integrates Packer with the QEMU plugin, standardizing the process across ISAs, as depicted in Figure~\ref{fig:disk-image-hypercalls}.
The base disk image creation workflow follows these steps:
\begin{enumerate}
    \item Automated Installation: Packer uses an official Ubuntu base ISO with Ubuntu Auto Install to set up a minimal server installation.
    \item gem5 Optimizations: The disk image is optimized for gem5 performance by:
    \begin{itemize}
        \item Installing the \texttt{gem5-bridge} driver and building \texttt{m5ops}.
        \item Disabling unnecessary services to improve boot times.
        These services include network and thermal management, which are not required for gem5 simulations.
    \end{itemize}
    \item Predefined Hypercalls: Three hypercalls (mechanisms for guest-host communication, detailed in Section~\ref{sec:hypercalls}) track key simulation stages:
    \begin{itemize}
        \item Kernel Boot Hypercall: Triggered when the kernel boots, before \texttt{systemd} starts.
        This hypercall is defined with the integer argument 1.
        \item Login Hypercall: Invoked after login but before the run script executes. For the base image, the run script is empty.
        This hypercall is defined with the integer argument 2.
        \item Exit Hypercall: Called after the run script completes, marking the end of the simulation.
        This hypercall is defined with the integer argument 3.
    \end{itemize}
    \item \textbf{Run Script Execution:} The run script, which is initiated after the login hypercall, facilitates the execution of a user-specified binary.
    This binary can be designated through two principal methods: either by embedding it within the disk image's creation pipeline or by providing it externally from the host system via the \texttt{set\_binary\_to\_run()}\footnote{\url{https://github.com/gem5/gem5/blob/v25.0.0.1/src/python/gem5/components/boards/kernel_disk_workload.py\#L263}} function.
\end{enumerate}

The key stages of disk image execution and hypercall triggers are visually represented in Figure~\ref{fig:hypercalls-boot-workflow}.


By default, the kernel boot and login hypercalls do not alter execution but serve as markers for users to monitor progress, while the exit hypercall exits the simulation, signalling its completion.
Additionally, all base disk images have their corresponding kernel builds as provided by Ubuntu available in gem5 resources.
These kernels include all the pre-installed modules along with the \texttt{gem5-bridge} driver as mentioned in Section~\ref{sec:gem5-bridge}.

Base disk images following this workflow are available for Ubuntu 22.04 LTS and 24.04 LTS across x86, ARM, and RISC-V ISAs, with the most recent kernel versions and \texttt{systemd} configurations.
To support faster development workflows, we also provide images without \texttt{systemd}, which remove startup overhead and enable quicker simulation cycles.
systemd-free images showed substantial boot-time reductions in our tested configurations (up to 23$\times$ on RISC-V, 22$\times$ on ARM, and 7$\times$ on x86).
This performance improvement makes the \texttt{systemd}-free images a good choice for rapid prototyping and regression testing.

Together, these images provide a standardized and flexible foundation for full-system simulations in gem5, significantly streamlining the disk image creation process.
Moreover, because our disk images are distributed through gem5 Resources and are decoupled from gem5's core simulation logic, they are not tied to any specific gem5 release.
The images remain valid across gem5 updates, with a field in their metadata indicating compatible gem5 versions.
This preserves long-term reproducibility without requiring disk image reconstruction.


\subsection{Extending Disk Images for Benchmark Suites}\label{sec:disk-image-extensions}
\begin{table}[h]
    \centering
    \caption{Benchmark and Disk Image Config Overview (x86, ARM, RISC-V)}
    \begin{tabular}{@{} >{\centering\arraybackslash}p{1.0cm} >{\centering\arraybackslash}p{1.8cm} p{2.5cm} >{\centering\arraybackslash}p{2cm} @{}}
    \toprule
    \textbf{Category} & \textbf{Benchmarks} & \textbf{Sizes/Graphs} & \textbf{Configs} \\ \midrule

    \multirow[c]{2}{*}{NPB} & \multirow[c]{2}{*}{\shortstack{ua, bt, cg, ep,\\ft, is, lu, mg, sp}} & \multirow[c]{2}{*}{\underline{Sizes}: S, A, B, C, D} & Ubuntu 24.04 LTS \\
                         &                                   &                        & (w \texttt{systemd}) \\ \midrule

    \multirow[c]{3}{*}{GAPBS} & \multirow[c]{3}{*}{\shortstack{bc, bfs, cc,\\pr, sssp, tc}} & \underline{Synthetic}: Test ($2^{19}$ nodes, deg 4), Train ($2^{21}$ nodes, deg 16), Set ($2^{24}$ nodes, deg 120) &  \\
                           &                            & \underline{Pre-Built}: soc-LiveJournal1~\cite{backstrom2006group, leskovec2009community}, facebook~\cite{leskovec2012learning} & \shortstack{Ubuntu 24.04 LTS \\ (w \texttt{systemd})} \\
                           &                            &                                                                                                           & \\ \midrule

    \multirow[c]{2}{*}{\shortstack{Base Ubuntu\\ 24.04 LTS}} & \multirow[c]{2}{*}{-} & \underline{Kernel}: 6.8.12/ & with \texttt{systemd} \\
                                                      &                    & 6.8.0-46-generic & w/o \texttt{systemd} \\ \midrule

    \multirow[c]{2}{*}{\shortstack{Base Ubuntu\\ 22.04 LTS}} & \multirow[c]{2}{*}{-} & \underline{Kernel}: 5.15.180/ & with \texttt{systemd} \\
                                                      &                    & 5.4.0-105-generic & w/o \texttt{systemd} \\ \bottomrule
    \end{tabular}
    \label{tab:disk-images}
\end{table}

The base images created using the workflow defined in Section~\ref{sec:disk-image-workflow-standardization} serve as a foundation for additional disk images containing annotated benchmarks to run in gem5's full-system mode.


We have extended the base images with scripts to install the NAS Parallel Benchmarks (NPB) and the GAP Benchmark Suite, with annotated hypercalls for regions of interest in each benchmark.
They are available for Ubuntu 24.04 LTS across x86, ARM, and RISC-V ISAs, with the most recent kernel versions and \texttt{systemd} configurations.
The disk images are optimized for gem5 performance and include the \texttt{gem5-bridge} driver, \texttt{m5ops}, and predefined hypercalls for monitoring simulation progress.
These images are enumerated in Table~\ref{tab:disk-images}, along with the benchmarks and dependencies like graphs available within each disk image.
The correctness, consistency, and reliability of the benchmarks and annotations within these extended disk images are thoroughly validated through comprehensive testing detailed in Section~\ref{sec:testing-validation}.



To illustrate the process of extending the base disk images with benchmarks, we present a case study using the STREAM memory bandwidth benchmark~\cite{mccalpin1995stream}.

\textit{Benchmark Preparation and Annotation}:
The STREAM benchmark is a simple synthetic benchmark that measures sustainable memory bandwidth and corresponding computation rates.
The preparation process follows these steps:
\begin{enumerate}
\item Source Acquisition: The benchmark source code was obtained from an implementation on GitHub~\cite{STREAM}.
\item Build System Modification: The makefile was updated to link against \texttt{libm5.a} to enable communication with gem5, add an \texttt{M5\_ANNOTATION} flag to conditionally include gem5-specific annotations, and ensure compilation with the \texttt{-no-pie} flag for compatibility with \texttt{m5ops}.
\item Hypercall Annotations: Hypercalls 4 and 5 were inserted before and after the main computational loop to mark the region of interest.
\end{enumerate}


\textit{Disk Image Creation}:
STREAM was installed on top of the base image using a file provisioner and a post-installation script that transferred the benchmark source code to the disk image and built the benchmark using \texttt{make}.

The file provisioner is used by Packer to copy the STREAM source code from the host machine to the disk image. For the STREAM benchmark disk image, the provisioner configuration is as follows:
\begin{lstlisting}[language=Python,
    basicstyle=\ttfamily\footnotesize,
    breaklines=true,
    breakatwhitespace=true
]
provisioner "file" {
    source      = "STREAM-master"
    destination = "/home/gem5/"
}
\end{lstlisting}

The resulting disk image includes the STREAM benchmark with properly annotated regions of interest.
It also inherits all optimizations and functionality from the base image, such as the \texttt{gem5-bridge} driver, \texttt{m5ops}, and predefined hypercalls, with just the addition of a few lines in the Packer configuration file.
Thus, our disk-image workflow improves flexibility and extensibility, enabling researchers to incorporate and annotate benchmarks for gem5 full-system simulations across all supported ISAs with minimal effort.
The code and configuration files used in this case study are available in the artifact repository.

\subsection{Testing and Validation}\label{sec:testing-validation}
To ensure the reliability and accuracy of the new disk images, we conducted extensive testing and validation.


Our testing process consisted of the following stages:
\begin{enumerate}
    \item Verifying Boot Cycle Hypercall Behavior: To confirm correct disk behavior, we check that the disk image boots successfully and the kernel boot hypercall, login hypercall, and exit hypercall are triggered at the appropriate stages.
    \item Ensuring Benchmark Hypercall Execution: We validate that benchmarks execute correctly and that the annotated hypercalls (hypercalls 4 and 5) are triggered as expected.
    \item Testing \texttt{gem5-bridge} Module: We verify that the \texttt{gem5-bridge} module functions as intended, allowing user-space execution of \texttt{m5} operations without root access.
    We do this by running a simple C binary in the disk image that maps the MMIO range and executes a simple exit hypercall.
    \item Verifying Workload Completion: We ensure that workloads complete successfully.
    This involves running a workload in gem5 and checking that the gem5 simulation completes without errors.
    In case of workloads with outputs like the NAS Parallel Benchmarks, we also verify that the gem5 simulation has the same expected output.
\end{enumerate}

For all tests, we used the prebuilt demo boards provided in the stable branch of the gem5 repository\footnote{\url{https://github.com/gem5/gem5/tree/v25.0.0.1/src/python/gem5/prebuilt/demo}}.
These demo boards cover the x86, ARM and RISC-V ISAs, and are configured with representative CPUs and memory systems.
Using standardized prebuilt boards ensures reproducibility and reduces setup complexity when evaluating features like hypercalls or workloads.

Through this testing process, we confirmed the correctness of our new features and identified a bug in gem5's ARM support.
When booting a Linux 6.8.12 kernel in gem5's TIMING mode on ARM, we observed that simulations would hang during early EL2 initialization when executing the \texttt{msr hcrx\_el2, x0} instruction.
Upon reaching the \texttt{msr} instruction, execution unexpectedly jumped to an invalid address (0x400) and subsequently attempted to execute an undefined instruction (0x000000), leading to an infinite fault loop at address 0x200.
We reported this issue to the gem5 community (Issue \#2116\footnote{\url{https://github.com/gem5/gem5/issues/2116}}).
With their help, the root cause was identified: the bootloader was not setting the \texttt{SCR\_EL3.HXEn} bit to 1, which is required for the kernel to boot in EL2.
We contributed a fix that explicitly sets this bit, ensuring the kernel boots into EL2 (hypervisor mode) as expected.
This patch has been merged into the gem5 repository and is available in the latest version\footnote{\url{https://github.com/gem5/gem5/pull/2130/}}. 




\subsection{Validation Results}\label{sec:validation-results}
This section presents the validation results confirming the functionality, integration, and reliability of the new features and resources we introduced, as described in the preceding sections.



\begin{figure}[htbp]
    \centering
    
\begin{lstlisting}[
basicstyle=\ttfamily\footnotesize,
breaklines=true,
breakatwhitespace=true
]
Dumping and resetting stats after kernel boot! Hypercall 1
Dumping and resetting stats after Ubuntu boot! Hypercall 2
Dumping and resetting stats at ROI begin! Hypercall 4
Dumping and resetting stats at ROI end! Hypercall 5
Dumping and resetting stats before exiting simulation! Hypercall 3
All exit events are called in expected order
Validation successful: BT (Class S) completed successfully.
\end{lstlisting}
    \caption{Sample output from the hypercall execution validator script, confirming that all hypercalls were executed at the expected stages during boot and benchmark execution.}
    \label{fig:hypercall-execution-validation}
\end{figure}

The end-to-end tests for the disk image creation workflow and the benchmark installation process also passed successfully.
The boot cycle hypercall behavior was verified across ISAs (Hypercalls 1, 2, and 3), confirming they were triggered at appropriate stages.
The benchmark hypercall execution was also verified (Hypercalls 4 and 5), confirming their correct placement and triggering within the application ROI.
Workloads tested completed successfully, and for those with defined outputs (like NPB), the gem5 simulation also produced the expected results.
A sample output from the hypercall execution validator script is shown in Figure~\ref{fig:hypercall-execution-validation}, confirming that all hypercalls were executed at the expected stages during boot and benchmark execution.

The GAPBS \texttt{tc} benchmark was excluded for the \texttt{soc-LiveJournal1} and \texttt{facebook} inputs because \texttt{tc} requires undirected graphs, while these datasets are directed.
For the larger workloads, i.e. the GAPBS \texttt{set} graph and the NPB Class~D and Class~F benchmarks, simulations were omitted due to memory constraints.
They have memory requirements exceeding 3~GiB, which is not supported on x86 systems in gem5, and require more than the 16~GiB allocated on ARM and RISC-V demo boards.

Furthermore, Table~\ref{tab:benchmark-stats} presents the simulated instruction counts within the defined regions of interest for the GAPBS and NPB benchmark suites, respectively, running on the new standardized disk images across ARM, x86, and RISC-V ISAs.
The consistency of these instruction counts across architectures provides strong quantitative evidence that the standardized benchmark disk images, the integrated \texttt{m5ops} and \texttt{gem5-bridge}, and the hypercall annotations are functioning correctly and provide a reproducible base for performance studies using gem5.
The disk image workflow guarantees software-level consistency, i.e., identical OS configuration, inputs, and execution paths leading up to the region of interest.
Microarchitectural state is intentionally left unconstrained to preserve meaningful cross-ISA comparisons.

\begin{table*}[htbp]
    \centering
    \caption{ROI Simulation Statistics for NPB and GAPBS Benchmarks across ARM, x86, and RISC-V ISAs. Sim~(s) is simulated seconds in the ROI. Host~(s) is wall-clock host seconds for the ROI. Insts~(Million) is the instruction count in millions within the ROI. LJ1 = soc-LiveJournal1.}
    \label{tab:benchmark-stats}
    \scriptsize
    \setlength{\tabcolsep}{3pt}
    \begin{tabular}{ll|>{\raggedleft\arraybackslash}p{0.9cm}>{\raggedleft\arraybackslash}p{1.3cm}>{\raggedleft\arraybackslash}p{1.3cm}|>{\raggedleft\arraybackslash}p{0.9cm}>{\raggedleft\arraybackslash}p{1.3cm}>{\raggedleft\arraybackslash}p{1.3cm}|>{\raggedleft\arraybackslash}p{0.9cm}>{\raggedleft\arraybackslash}p{1.3cm}>{\raggedleft\arraybackslash}p{1.3cm}}
        \toprule
        & & \multicolumn{3}{c|}{\textbf{ARM}} & \multicolumn{3}{c|}{\textbf{x86}} & \multicolumn{3}{c}{\textbf{RISC-V}} \\
        \textbf{BMark} & \textbf{Size} & \textbf{Sim (s)} & \textbf{Host (s)} & \textbf{Insts (M)} & \textbf{Sim (s)} & \textbf{Host (s)} & \textbf{Insts (M)} & \textbf{Sim (s)} & \textbf{Host (s)} & \textbf{Insts (M)} \\
        \midrule
        \multicolumn{11}{c}{\textit{NAS Parallel Benchmarks (NPB)}} \\
        \midrule
        BT & S & 0.09 & 9.27 $\times$ 10\textsuperscript{2} & 323.86 & 0.18 & 1.29 $\times$ 10\textsuperscript{3} & 494.78 & 0.28 & 9.66 $\times$ 10\textsuperscript{2} & 400.13 \\
           & A & 80.41 & 6.93 $\times$ 10\textsuperscript{5} & 230377.06 & 148.81 & 7.94 $\times$ 10\textsuperscript{5} & 356866.41 & 216.35 & 7.17 $\times$ 10\textsuperscript{5} & 283947.66 \\
           & B & 351.28 & 2.97 $\times$ 10\textsuperscript{6} & 956813.47 & 636.72 & 3.38 $\times$ 10\textsuperscript{6} & 1482418.91 & 915.29 & 2.93 $\times$ 10\textsuperscript{6} & 1179935.47 \\
        \cmidrule(lr){1-11}
        CG & S & 0.06 & 6.04 $\times$ 10\textsuperscript{2} & 268.34 & 0.08 & 5.23 $\times$ 10\textsuperscript{2} & 271.91 & 0.17 & 1.03 $\times$ 10\textsuperscript{3} & 343.43 \\
           & A & 2.19 & 1.76 $\times$ 10\textsuperscript{4} & 5956.23 & 2.59 & 1.51 $\times$ 10\textsuperscript{4} & 6059.46 & 5.12 & 1.64 $\times$ 10\textsuperscript{4} & 7534.78 \\
           & B & 104.60 & 7.08 $\times$ 10\textsuperscript{5} & 218548.11 & 129.22 & 6.51 $\times$ 10\textsuperscript{5} & 220714.84 & 280.04 & 6.81 $\times$ 10\textsuperscript{5} & 275534.13 \\
        \cmidrule(lr){1-11}
        EP & S & 0.27 & 3.04 $\times$ 10\textsuperscript{3} & 1399.93 & 0.52 & 3.32 $\times$ 10\textsuperscript{3} & 1888.39 & 0.85 & 2.99 $\times$ 10\textsuperscript{3} & 1520.00 \\
           & A & 4.30 & 4.66 $\times$ 10\textsuperscript{4} & 22356.14 & 8.26 & 5.25 $\times$ 10\textsuperscript{4} & 30224.48 & 13.57 & 4.82 $\times$ 10\textsuperscript{4} & 24270.32 \\
           & B & 17.16 & 1.84 $\times$ 10\textsuperscript{5} & 89374.21 & 33.02 & 2.11 $\times$ 10\textsuperscript{5} & 120833.41 & 54.28 & 1.95 $\times$ 10\textsuperscript{5} & 97044.90 \\
        \cmidrule(lr){1-11}
        FT & S & 0.11 & 1.14 $\times$ 10\textsuperscript{3} & 417.08 & 0.19 & 1.20 $\times$ 10\textsuperscript{3} & 580.25 & 0.29 & 1.93 $\times$ 10\textsuperscript{3} & 498.15 \\
           & A & 4.27 & 4.98 $\times$ 10\textsuperscript{4} & 16229.69 & 7.38 & 5.32 $\times$ 10\textsuperscript{4} & 22716.86 & 12.60 & 5.01 $\times$ 10\textsuperscript{4} & 19208.10 \\
           & B & 53.84 & 6.29 $\times$ 10\textsuperscript{5} & 205541.45 & 92.82 & 6.42 $\times$ 10\textsuperscript{5} & 285279.66 & 163.39 & 6.63 $\times$ 10\textsuperscript{5} & 239981.83 \\
        \cmidrule(lr){1-11}
        IS & S & 0.01 & 3.71 $\times$ 10\textsuperscript{1} & 18.85 & 0.01 & 3.46 $\times$ 10\textsuperscript{1} & 16.40 & 0.01 & 6.97 $\times$ 10\textsuperscript{1} & 21.65 \\
           & A & 0.78 & 4.36 $\times$ 10\textsuperscript{3} & 2303.37 & 1.04 & 4.65 $\times$ 10\textsuperscript{3} & 1983.13 & 2.00 & 7.88 $\times$ 10\textsuperscript{3} & 2646.83 \\
           & B & 3.09 & 1.77 $\times$ 10\textsuperscript{4} & 9220.36 & 4.16 & 1.85 $\times$ 10\textsuperscript{4} & 7923.25 & 8.02 & 2.13 $\times$ 10\textsuperscript{4} & 10578.77 \\
        \cmidrule(lr){1-11}
        LU & S & 0.03 & 3.12 $\times$ 10\textsuperscript{2} & 120.31 & 0.06 & 3.42 $\times$ 10\textsuperscript{2} & 157.08 & 0.10 & 5.61 $\times$ 10\textsuperscript{2} & 147.84 \\
           & A & 42.31 & 3.27 $\times$ 10\textsuperscript{5} & 111701.98 & 71.50 & 3.46 $\times$ 10\textsuperscript{5} & 153606.93 & 117.34 & 3.69 $\times$ 10\textsuperscript{5} & 150098.27 \\
           & B & 185.42 & 1.43 $\times$ 10\textsuperscript{6} & 457908.24 & 305.72 & 1.50 $\times$ 10\textsuperscript{6} & 629667.56 & 498.03 & 1.59 $\times$ 10\textsuperscript{6} & 620347.77 \\
        \cmidrule(lr){1-11}
        MG & S & 0.00 & 3.92 $\times$ 10\textsuperscript{1} & 15.11 & 0.01 & 3.75 $\times$ 10\textsuperscript{1} & 16.78 & 0.02 & 6.40 $\times$ 10\textsuperscript{1} & 30.08 \\
           & A & 1.64 & 1.87 $\times$ 10\textsuperscript{4} & 5705.09 & 2.38 & 1.61 $\times$ 10\textsuperscript{4} & 6099.56 & 8.59 & 3.09 $\times$ 10\textsuperscript{4} & 13093.21 \\
           & B & 7.61 & 8.70 $\times$ 10\textsuperscript{4} & 26300.27 & 11.08 & 7.42 $\times$ 10\textsuperscript{4} & 28226.68 & 40.29 & 1.48 $\times$ 10\textsuperscript{5} & 61472.02 \\
        \cmidrule(lr){1-11}
        SP & S & 0.04 & 4.16 $\times$ 10\textsuperscript{2} & 156.15 & 0.07 & 4.00 $\times$ 10\textsuperscript{2} & 187.45 & 0.13 & 4.46 $\times$ 10\textsuperscript{2} & 198.78 \\
           & A & 58.15 & 3.71 $\times$ 10\textsuperscript{5} & 122110.07 & 78.22 & 3.35 $\times$ 10\textsuperscript{5} & 150770.66 & 123.48 & 3.93 $\times$ 10\textsuperscript{5} & 156482.19 \\
           & B & 248.64 & 1.52 $\times$ 10\textsuperscript{6} & 489937.47 & 328.07 & 1.36 $\times$ 10\textsuperscript{6} & 611865.63 & 499.82 & 1.59 $\times$ 10\textsuperscript{6} & 649768.40 \\
        \cmidrule(lr){1-11}
        UA & S & 0.88 & 8.24 $\times$ 10\textsuperscript{3} & 3491.66 & 1.26 & 7.93 $\times$ 10\textsuperscript{3} & 3434.25 & 3.17 & 1.04 $\times$ 10\textsuperscript{4} & 4762.60 \\
           & A & 58.29 & 3.80 $\times$ 10\textsuperscript{5} & 158967.09 & 71.20 & 3.64 $\times$ 10\textsuperscript{5} & 156615.62 & 147.97 & 4.65 $\times$ 10\textsuperscript{5} & 210103.77 \\
           & B & 277.46 & 1.57 $\times$ 10\textsuperscript{6} & 632845.57 & 316.77 & 1.49 $\times$ 10\textsuperscript{6} & 633035.69 & 591.94 & 1.80 $\times$ 10\textsuperscript{6} & 850049.58 \\
        \midrule
        \multicolumn{11}{c}{\textit{GAP Benchmark Suite (GAPBS)}} \\
        \midrule
        \textbf{BMark} & \textbf{Graph} & \textbf{Sim (s)} & \textbf{Host (s)} & \textbf{Insts (M)} & \textbf{Sim (s)} & \textbf{Host (s)} & \textbf{Insts (M)} & \textbf{Sim (s)} & \textbf{Host (s)} & \textbf{Insts (M)} \\
        \midrule
        BC & test & 0.18 & 4.96 $\times$ 10\textsuperscript{2} & 195.15 & 0.21 & 5.11 $\times$ 10\textsuperscript{2} & 184.96 & 0.59 & 5.97 $\times$ 10\textsuperscript{2} & 194.87 \\
           & train & 3.42 & 5.96 $\times$ 10\textsuperscript{3} & 2439.73 & 3.98 & 6.48 $\times$ 10\textsuperscript{3} & 2374.57 & 7.37 & 7.09 $\times$ 10\textsuperscript{3} & 2388.21 \\
           & Facebook & 0.00 & 4.04 $\times$ 10\textsuperscript{0} & 2.34 & 0.00 & 4.46 $\times$ 10\textsuperscript{0} & 2.59 & 0.00 & 4.48 $\times$ 10\textsuperscript{0} & 2.46 \\
           & LJ1 & 2.69 & 5.72 $\times$ 10\textsuperscript{3} & 2528.60 & 3.11 & 6.13 $\times$ 10\textsuperscript{3} & 2541.57 & 5.39 & 6.73 $\times$ 10\textsuperscript{3} & 2600.90 \\
        \cmidrule(lr){1-11}
        BFS & test & 0.02 & 1.31 $\times$ 10\textsuperscript{2} & 67.70 & 0.02 & 1.37 $\times$ 10\textsuperscript{2} & 73.33 & 0.05 & 1.55 $\times$ 10\textsuperscript{2} & 75.82 \\
            & train & 0.10 & 3.46 $\times$ 10\textsuperscript{2} & 168.58 & 0.11 & 3.69 $\times$ 10\textsuperscript{2} & 178.97 & 0.20 & 3.96 $\times$ 10\textsuperscript{2} & 175.27 \\
            & Facebook & 0.00 & 2.72 $\times$ 10\textsuperscript{0} & 1.65 & 0.00 & 6.24 $\times$ 10\textsuperscript{0} & 2.13 & 0.00 & 3.50 $\times$ 10\textsuperscript{0} & 1.94 \\
            & LJ1 & 0.22 & 1.24 $\times$ 10\textsuperscript{3} & 647.75 & 0.25 & 1.30 $\times$ 10\textsuperscript{3} & 701.24 & 0.53 & 1.44 $\times$ 10\textsuperscript{3} & 713.83 \\
        \cmidrule(lr){1-11}
        CC & test & 0.03 & 1.94 $\times$ 10\textsuperscript{2} & 94.52 & 0.03 & 1.87 $\times$ 10\textsuperscript{2} & 97.02 & 0.08 & 2.03 $\times$ 10\textsuperscript{2} & 92.91 \\
           & train & 0.15 & 8.70 $\times$ 10\textsuperscript{2} & 387.06 & 0.16 & 8.42 $\times$ 10\textsuperscript{2} & 395.90 & 0.39 & 9.09 $\times$ 10\textsuperscript{2} & 390.67 \\
           & Facebook & 0.00 & 5.97 $\times$ 10\textsuperscript{0} & 3.71 & 0.00 & 6.66 $\times$ 10\textsuperscript{0} & 4.26 & 0.00 & 8.01 $\times$ 10\textsuperscript{0} & 4.42 \\
           & LJ1 & 0.29 & 1.32 $\times$ 10\textsuperscript{3} & 650.00 & 0.31 & 1.26 $\times$ 10\textsuperscript{3} & 652.19 & 0.59 & 1.33 $\times$ 10\textsuperscript{3} & 623.47 \\
        \cmidrule(lr){1-11}
        CC\_ & test & 0.07 & 4.12 $\times$ 10\textsuperscript{2} & 171.64 & 0.09 & 4.39 $\times$ 10\textsuperscript{2} & 186.29 & 0.44 & 5.43 $\times$ 10\textsuperscript{2} & 193.13 \\
        SV   & train & 1.95 & 5.55 $\times$ 10\textsuperscript{3} & 2244.02 & 2.61 & 6.96 $\times$ 10\textsuperscript{3} & 2388.45 & 10.13 & 8.75 $\times$ 10\textsuperscript{3} & 2575.31 \\
              & Facebook & 0.00 & 5.51 $\times$ 10\textsuperscript{0} & 1.86 & 0.00 & 4.02 $\times$ 10\textsuperscript{0} & 2.40 & 0.00 & 4.03 $\times$ 10\textsuperscript{0} & 2.17 \\
              & LJ1 & 2.49 & 4.60 $\times$ 10\textsuperscript{3} & 2088.72 & 2.99 & 4.82 $\times$ 10\textsuperscript{3} & 2169.90 & 7.92 & 6.06 $\times$ 10\textsuperscript{3} & 2350.27 \\
        \cmidrule(lr){1-11}
        PR & test & 0.12 & 7.94 $\times$ 10\textsuperscript{2} & 237.75 & 0.15 & 7.45 $\times$ 10\textsuperscript{2} & 247.68 & 0.89 & 9.16 $\times$ 10\textsuperscript{2} & 308.87 \\
           & train & 2.22 & 5.82 $\times$ 10\textsuperscript{3} & 1571.65 & 2.91 & 6.40 $\times$ 10\textsuperscript{3} & 1618.30 & 11.96 & 8.51 $\times$ 10\textsuperscript{3} & 2205.01 \\
           & Facebook & 0.00 & 4.39 $\times$ 10\textsuperscript{0} & 2.41 & 0.00 & 4.87 $\times$ 10\textsuperscript{0} & 2.91 & 0.00 & 5.04 $\times$ 10\textsuperscript{0} & 3.00 \\
           & LJ1 & 9.52 & 2.80 $\times$ 10\textsuperscript{4} & 9145.10 & 11.38 & 2.71 $\times$ 10\textsuperscript{4} & 9466.30 & 29.33 & 3.16 $\times$ 10\textsuperscript{4} & 12222.92 \\
        \cmidrule(lr){1-11}
        PR\_ & test & 0.21 & 1.39 $\times$ 10\textsuperscript{3} & 423.28 & 0.26 & 1.32 $\times$ 10\textsuperscript{3} & 439.13 & 1.59 & 1.75 $\times$ 10\textsuperscript{3} & 544.06 \\
        SPMV & train & 3.37 & 8.87 $\times$ 10\textsuperscript{3} & 2373.51 & 4.32 & 9.60 $\times$ 10\textsuperscript{3} & 2442.11 & 18.39 & 1.38 $\times$ 10\textsuperscript{4} & 3319.20 \\
              & Facebook & 0.00 & 3.01 $\times$ 10\textsuperscript{1} & 16.51 & 0.00 & 2.58 $\times$ 10\textsuperscript{1} & 16.06 & 0.01 & 4.19 $\times$ 10\textsuperscript{1} & 23.22 \\
              & LJ1 & 9.43 & 2.86 $\times$ 10\textsuperscript{4} & 9490.58 & 11.22 & 2.75 $\times$ 10\textsuperscript{4} & 9799.82 & 31.07 & 3.77 $\times$ 10\textsuperscript{4} & 12563.18 \\
        \cmidrule(lr){1-11}
        SSSP & test & 0.09 & 3.07 $\times$ 10\textsuperscript{2} & 130.80 & 0.11 & 3.58 $\times$ 10\textsuperscript{2} & 138.17 & 0.34 & 4.34 $\times$ 10\textsuperscript{2} & 145.03 \\
             & train & 1.57 & 3.43 $\times$ 10\textsuperscript{3} & 1307.57 & 1.98 & 4.37 $\times$ 10\textsuperscript{3} & 1448.42 & 4.50 & 5.05 $\times$ 10\textsuperscript{3} & 1474.03 \\
             & Facebook & 0.00 & 1.04 $\times$ 10\textsuperscript{1} & 6.20 & 0.00 & 1.20 $\times$ 10\textsuperscript{1} & 6.81 & 0.00 & 1.46 $\times$ 10\textsuperscript{1} & 7.62 \\
             & LJ1 & 1.78 & 3.78 $\times$ 10\textsuperscript{3} & 1487.08 & 2.15 & 4.55 $\times$ 10\textsuperscript{3} & 1618.80 & 4.26 & 5.17 $\times$ 10\textsuperscript{3} & 1635.94 \\
        \cmidrule(lr){1-11}
        TC & test & 0.08 & 2.46 $\times$ 10\textsuperscript{2} & 132.96 & 0.10 & 3.10 $\times$ 10\textsuperscript{2} & 166.39 & 0.25 & 2.86 $\times$ 10\textsuperscript{2} & 113.21 \\
           & train & 3.99 & 1.00 $\times$ 10\textsuperscript{4} & 5819.92 & 4.60 & 1.20 $\times$ 10\textsuperscript{4} & 7461.71 & 7.24 & 1.04 $\times$ 10\textsuperscript{4} & 4997.11 \\
           & Facebook & --- & --- & --- & --- & --- & --- & --- & --- & --- \\
           & LJ1 & --- & --- & --- & --- & --- & --- & --- & --- & --- \\
        \bottomrule
    \end{tabular}
\end{table*}

\section{Decoupled Guest-Host Communication}\label{sec:guest-host-communication}
We update gem5's exit event system and introduce hypercalls to address the tight coupling and lack of context in the classic generator-based exit event system.

\subsection{Hypercalls}\label{sec:hypercalls}
We updated the exit event system to a more flexible class-based model, moving away from the older generator-based approach.
This change decouples exit event handling from workload behavior, allowing users to handle exit events in any order, independent of how they are triggered by the workload.

To address the lack of simulation context in exit events, we introduced the \texttt{m5 hypercall} command as part of the \texttt{m5ops} library.
The name \textit{hypercalls} originates from virtualized systems, where a guest OS uses them to communicate with the host OS.
Similarly, in gem5, hypercalls are a mechanism for the guest application (or OS) running inside the simulation to communicate with the host simulator.

Using the syntax \texttt{m5 hypercall <num>}, a program running inside the simulation can send an integer value (\texttt{<num>}) to the host.
How the host interprets this integer and what actions it triggers is defined by the user within their simulation script.
For instance, receiving the integer \texttt{4} via \texttt{m5 hypercall 4} could be programmed in the host script to signify the start of a benchmark's region of interest, triggering a statistics dump and reset.
This flexibility enables dynamic control over simulations, supporting tasks like on-the-fly parameter changes, checkpoint restoration, or annotating specific phases within a running workload.
When a hypercall pseudo-instruction is received, it pauses the simulation, executes the corresponding host logic, and then resumes without advancing the simulated cycles.
Therefore, they do not introduce additional simulated time nor cause synchronization issues.

Additionally, the new class-based structure allows gem5 to support hypercalls triggered from outside the simulated system.
For example, a user can invoke a hypercall directly from the gem5 configuration script and optionally attach a custom payload.
This payload, provided as a Python dictionary, allows further customization of the hypercall's behavior based on its contents.
This feature enables external control over the simulation, allowing researchers to make their own custom tools that interact with running gem5 simulations.
For example, we have implemented an external signal utility (Section~\ref{sec:hypercall-external-signal}) that allows the users to remotely monitor running gem5 simulations.


\subsection{Hypercall External Signal Utility}\label{sec:hypercall-external-signal}
We also implemented the hypercall external signal utility, which allows users to remotely trigger a hypercall during simulation execution.
This utility is particularly useful for monitoring simulation progress, as it can return the current status and statistics of the simulation at the point where the utility is invoked.

The hypercall mechanism relies on inter-process communication via Unix sockets and shared memory.
We implemented a dedicated utility script to send hypercalls to a target gem5 simulation.
This script sends JSON-formatted payloads through shared memory and trigger a \texttt{SIGHUP} signal to the gem5 process, whose PID is accepted via the command line, which is configured to handle and respond to such signals.

Additionally, the utility can send snapshots of gem5 statistics, enabling users to track simulation progress remotely.
It also enables dynamic control of debug flags during runtime, allowing users to adjust instrumentation levels without restarting simulations.
These capabilities provide flexibility for monitoring and debugging complex simulation workflows.
A sample usage of the hypercall external signal utility is shown in Figure~\ref{fig:hypercall-external-signal-usage}.

\begin{figure}[htbp]
    \centering
\begin{lstlisting}[
basicstyle=\ttfamily\scriptsize,
breaklines=true,
breakatwhitespace=true,
columns=flexible
]
% python3 ./hypercall_external_signal/orchestrator-request.py
  --pid 3605886 status
Response: {"workload": "x86-ubuntu-24.04-npb-bt-s",
  "tick": 368555630112, "sim_id": null,
  "curr_instructions_executed": 658201400}

% python3 ./hypercall_external_signal/orchestrator-request.py
  --pid 3605886 status
Response: {"workload": "x86-ubuntu-24.04-npb-bt-s",
  "tick": 379428293898, "sim_id": null,
  "curr_instructions_executed": 683297124}
\end{lstlisting}    \caption{Example usage of the hypercall external signal utility to fetch simulation statistics from a running gem5 process.}
    \label{fig:hypercall-external-signal-usage}
\end{figure}




\section{Parallel Experiment Orchestration}\label{sec:multisim-suite}

We introduced two features to address the challenge of running multiple workloads within a single simulation: Suite and MultiSim.
Together, they are intended to automate orchestration of experiments in parallel and eliminate most manual setup or external scripting.

The Suite class provides a structured way to group related simulation resources, such as benchmarks, workloads, and disk images, mirroring the concept of traditional benchmark suites used in computer architecture research (e.g., NAS Parallel Benchmarks~\cite{npb}, GAP Benchmark Suite~\cite{beamer2015gap})~\cite{gem5_suites_real}.
Each component of a Suite has a unique ID and configuration, and can be easily iterated over, filtered, or executed in specific subsets.
This simplifies the setup for experiments involving large numbers of predefined configurations.

Complementing Suite is MultiSim, which provides functionality for launching multiple gem5 simulations concurrently.
This feature is intended for scenarios like running multiple workloads that are part of the same suite or evaluating design variations by sweeping parameters (e.g., cache sizes, associativity).
Instead of manually launching and managing individual simulation runs with Bash scripts, MultiSim handles the orchestration internally.
Under the hood, MultiSim constructs a list of simulator objects, each with a unique ID, and executes them across a user-specified number of processes, running each simulation instance until all are complete.
The intended benefit of this approach is to \textit{reduce external Bash scripting} previously needed to manage multiple independent gem5 runs.


The synergy between Suite and MultiSim is evident when they are combined.
gem5 users can define a Suite containing numerous benchmarks or configurations and then use MultiSim to execute them concurrently, with far less manual intervention and reducing overall workflow complexity.

To demonstrate the practical advantages, particularly in workflow simplification, we conducted a comparative experiment:

\begin{itemize}
    \item \textit{Baseline}: Executed the Vertical Research Group's microbenchmark suite~\cite{nowatzki2015architectural} on gem5's prebuilt RISC-V demo board\footnote{\url{https://github.com/gem5/gem5/blob/v25.0.0.1/src/python/gem5/prebuilt/demo/riscv_demo_board.py}} in SE mode.
    Benchmarks were run sequentially via Bash script.
    \item \textit{Suite + Bash}: Executed the same microbenchmarks using the Suite feature and the prebuilt RISC-V demo board.
    Benchmarks were run sequentially via Bash script.
    \item \textit{MultiSim + Suite}: Executed the Suite using MultiSim for concurrent runs, testing up to 8 and 33 simulations.
\end{itemize}

While concurrency inherently offers performance benefits, the primary focus here is the impact on workflow complexity, measured by the lines of configuration code required.



The Baseline approach required 70 lines of Bash scripting and 33 lines of Python configuration to orchestrate and run the experiments.
The Suite + Bash approach showed similar complexity (72 lines of Bash, 41 lines of Python) since it still required external scripting for sequential execution.
However, combining MultiSim with Suite significantly simplified the setup, eliminating the need for Bash scripting \textit{entirely} and reducing the Python configuration to just \textit{16 lines}.
The following code snippet shows the configuration needed to run all 33 microbenchmarks concurrently using MultiSim and Suite (excluding imports):

\begin{lstlisting}[
    language=Python,
    basicstyle=\ttfamily\footnotesize,
    breaklines=true,
    breakatwhitespace=true
]
multisim.set_num_processes(33)

for workload in obtain_resource("riscv-vertical-microbenchmarks"):
    board = RiscvDemoBoard()
    board.set_workload(workload)
    multisim.add_simulator(
        Simulator(board=board, id=f"process_{workload.get_id()}")
    )
\end{lstlisting}

This simplification comes from MultiSim handling a substantial portion of the simulation management and orchestration tasks that were previously implemented through external scripts.



In summary, the Suite and MultiSim features in gem5 provide a mechanism for automating and streamlining experiments involving multiple configurations or workloads.
Their primary benefit is reducing manual orchestration, particularly reliance on complex Bash scripts, which helps make workflows less error-prone, improves reproducibility, and lowers the barrier to conducting large-scale studies.
Overall, their core value lies in improving ease-of-use and experimental management.

\section{Discussion}\label{sec:discussion}

\subsection{Design Tradeoffs}
\textbf{MultiSim's process-based concurrency} provides isolation between simulations, preventing errors in one experiment from affecting others.
Although this initially appears restrictive, it encourages cleaner separation: checkpoint triggers become part of the workload specification rather than external logic, making experiments more self-contained and portable.
We tested up to 33 concurrent simulations; practical limits depend on host resources (CPU cores, memory, disk I/O), as each gem5 instance consumes significant memory.

\textbf{Suite's static workload model} prioritizes repeatability. 
Once defined, a Suite contains a fixed collection of workloads that cannot change during execution. 
This design ensures that citing ``NPB Suite v1.0'' always refers to the same set of benchmarks, making experiments reproducible across time and research groups.

\textbf{Standardized disk images} target common research scenarios.
We focus on Ubuntu Linux as it covers the majority of architectural studies.
Rather than enforcing a single uniform configuration, we provide composable building blocks that are reproducible individually and freely combinable.
A researcher studying an unconventional memory hierarchy or a specialized operating system, for example, can reuse our validated base image and hypercall infrastructure while adding their own provisioning layers on top via Packer, without touching any of gem5's core simulation logic.

\subsection{Current Limitations}
\textbf{Heterogeneous systems:} Our workflow targets homogeneous CPUs.
Extending to CPU+GPU or custom accelerators requires specialized device drivers and boot sequences that do not fit neatly into our standardized approach.

\textbf{x86 memory constraints:} The x86 demo-board setup is limited to 3 GiB of guest physical memory, so larger benchmark configurations may not fit without board/address-space changes.
Resolving this requires non-trivial changes to gem5's address space management, which is beyond our current scope.

\textbf{Accuracy:} While we validate instruction count consistency, we do not claim cycle-accurate performance relative to hardware.
Our contribution is \textit{reproducible} simulation, not necessarily \textit{accurate} performance prediction.

\subsection{Enabling New Research Paradigms}
Our contributions enable new research paradigms previously impractical with gem5.
MultiSim's concurrency allows large-scale parameter sweeps across hundreds of configurations, enabling design space exploration studies that were computationally prohibitive with sequential execution~\cite{ipek2008dse,shao2014aladdin}.
Standardized resources facilitate cross-ISA comparisons, where researchers can now easily reproduce and extend work across x86, ARM, and RISC-V platforms without rebuilding experimental infrastructure.
Hypercalls support adaptive simulations, such as runtime statistics collection for long-running workloads, opening avenues for workload characterization and phase-based analysis~\cite{calzarossa2016workload,sherwood2004phases}.
Researchers can also efficiently bundle and share their own resources and workloads as suites, providing simple scripts that automatically iterate through and run all benchmarks, providing gem5 support for new workloads out of the box.
By reducing setup complexity and eliminating the need for custom scripting, these tools lower barriers to entry for new researchers while enabling experienced users to conduct more sophisticated architectural studies.

\section{Related Work}\label{sec:related-work}

\textbf{Reproducibility frameworks:} Reproducibility remains a major challenge in research, motivating frameworks that enhance transparency and repeatability.
The Collective Knowledge (CK) initiative~\cite{ck} addresses this by organizing experiments as reusable components and portable workflows with common APIs.
CK supports automated benchmarking, tuning, and artifact evaluation, and its \href{cKnowledge.io}{cKnowledge.io} platform enables sharing and management of these workflows.
Other tools like Docker~\cite{merkel2014docker} for containerization, and workflow managers such as Snakemake~\cite{molder2025sustainable} and Nextflow~\cite{di2017nextflow}, also support reproducible computational pipelines, though they are not specific to architectural simulation.
Similarly, gem5 Resources is a centralized platform for sharing and discovering simulation assets.

\textbf{gem5 ecosystem:} Within the gem5 community, there have been several efforts to improve reproducibility and ease-of-use.
The gem5 Vision project~\cite{gem5vision2023poster} introduced a web interface for gem5 Resources\footnote{\url{https://resources.gem5.org/}}, creating a centralized platform for discovering and downloading resources.
Additionally, the \texttt{obtain\_resource} function was updated to support a MongoDB database, streamlining resource retrieval and ensuring compatibility with different gem5 versions.
A similar effort to improve the reproducibility of gem5 simulations is the gem5ART project~\cite{bruce2021enabling}, which is a framework for managing and executing gem5 simulations in a reproducible manner.
The gem5ART project leveraged databases to store results from gem5 simulations, enabling users to easily compare and reproduce experiments.
The gem5ART project also introduced the concept of ``artifacts'' to encapsulate the resources and dependencies required to run a simulation, similar to the concept of workloads in gem5 Resources.
Our work builds on these initiatives by introducing features that further improve the reproducibility and standardization of gem5 simulations.

\section{Conclusion}\label{sec:conclusion}
Our contributions strengthen reproducibility in gem5 by standardizing the artifacts and workflows researchers depend on.
The updated Packer-based disk-image workflow, combined with validated NPB and GAPBS images across x86, ARM, and RISC-V, establishes a consistent baseline for full-system studies and standardizes the runtime environment to reduce result perturbations.
Importantly, this standardization acts as a composable foundation rather than a uniform constraint, i.e., researchers can extend these building blocks for conventional and unconventional experiments, while benefiting from validated, reproducible baselines.
Hypercalls decouple guest-host communication and make simulation control easier to customize.
Suites and MultiSim remove much of the external scripting previously required to run large sets of workloads and parallelize instances of gem5, resulting in simpler experiment orchestration.
These changes reduce setup overhead, standardize execution, and make it easier for researchers to run, share, and extend gem5-based studies with consistent results.

\section*{Acknowledgments}
We thank the anonymous reviewers for their thoughtful feedback and suggestions, which helped improve the clarity and presentation of this work.
We also thank the gem5 community for their support and feedback during the development of these features, particularly Giacomo Travaglini for his help in debugging the ARM EL2 boot issue.
We are also grateful to the members of the Davis Architecture Research Group (DArchR) for their valuable feedback and support throughout this project.
This work was supported by the National Science Foundation (NSF) under Grant 2311888, by a grant from Los Alamos National Laboratory, and by the U.S. Department of Energy under Award SC0024502.

\bibliography{references}

\bibliographystyle{IEEEtranS}

\appendices
\section{Artifact Evaluation Appendix}

\subsection{Abstract}

This paper introduces standardized gem5 resources (disk images, kernels and workloads) and new features (hypercalls, Suite, MultiSim) to enhance reproducibility in gem5-based research.
All the contributions of this paper are upstreamed to the main gem5 repository and are available in release v25.0 and later.
The source for building all the resources mentioned in this paper is available in the gem5 resources repository. For ease of access, we have also created a Zenodo archive with all the sources and scripts needed to run a representative simulation in gem5.

Since the paper includes building multiple disk images and workloads across different ISAs and running gem5 simulations to validate them, a process that would take weeks in total, these instructions focus on rebuilding the X86 Ubuntu 24.04 NPB disk image from the provided base image and running the NPB BT Size S workload as an example.

\subsection{Artifact check-list (meta-information)}

{\small
\begin{itemize}
  \item {\bf Program:} gem5 v25.0, Packer, QEMU
  \item {\bf Compilation:} gem5 built from source using SCons;
        disk images built using Packer with QEMU plugin
  \item {\bf Binary:} Pre-built x86 base disk image provided in archive (used only for building the NPB disk image); NPB disk image built locally by reviewer using provided Packer scripts and used for simulation
  \item {\bf Run-time environment:} Ubuntu 22.04 or 24.04 host;
        KVM required for x86 disk image builds
  \item {\bf Hardware:} x86 host machine with KVM support
  \item {\bf Execution:} Packer for disk image build; gem5
        full-system simulation (single run for evaluation)
  \item {\bf Metrics:} ROI instruction count
        (\texttt{ROIsimInsts}) extracted from gem5 stats file
  \item {\bf Output:} Hypercall execution log matching Figure~3;
        ROI instruction count matching Figure~5
  \item {\bf Experiments:} Single NPB BT Size~S simulation on x86
        for evaluation; full reproduction of Table~2
        requires all ISAs and benchmark configurations and is
        estimated to take weeks to months.
  \item {\bf How much disk space required (approximately)?:}
        $\sim$~10GB for disk images and gem5 build
  \item {\bf How much time is needed to complete experiments
        (approximately)?:} $\sim$20 minutes to build gem5 $+ \sim$10 minutes to build the disk images $+ \sim$6 hours (background) for NPB BT Size~S simulation run
  \item {\bf Publicly available?:} Yes
  \item {\bf Code licenses (if publicly available)?:} BSD-3-Clause
  \item {\bf Archived (provide DOI)?:} \texttt{10.5281/zenodo.18912932} ~\cite{pai2026gem5artifacts}
\end{itemize}
}

\subsection{Description}

\subsubsection{How to access}

Everything needed to reproduce the evaluation is contained in the
Zenodo archive (DOI:~\texttt{10.5281/zenodo.18912932}). The archive includes:
\begin{itemize}
  \item Packer scripts and build scripts for the base Ubuntu x86,
        ARM, and RISC-V disk images and all benchmark images
        (NPB, GAPBS) described in the paper
  \item A gem5 simulation script for running NPB benchmarks
  \item A stats extraction script used for parsing the gem5 stats file and extracting the ROI instruction count metric.
  \item A \texttt{create\_resources.py} script that generates a resources.json file with the correct paths and checksums for the built disk image, kernel, and a workload to run the NPB BT Class S simulation.
  \item Source for gem5 version 25.0, which is the version used for evaluation.
\end{itemize}

The upstream versions of the disk image sources are maintained in
the gem5-resources repository.\footnote{\url{https://github.com/gem5/gem5-resources}}

Note: The NPB Packer script in the Zenodo archive is modified from the upstream version to accept a local base image path instead of downloading one from gem5 Resources. No other changes were made. The resulting disk image is identical.

\subsubsection{Hardware dependencies}

An x86 host machine with KVM support is required for building x86 disk images.
RISC-V images are built without KVM and the build process is slower.

\subsubsection{Software dependencies}

\begin{itemize}
  \item Packer, which is downloaded automatically by the
        \texttt{build-*.sh} scripts, present in each disk image source directory.
  \item QEMU with KVM support \url{https://www.qemu.org/download/#linux}
  \item gem5 v25.0 build dependencies: GCC$>$=10, Python3.6+, SCons
        (see gem5 build documentation at
        \url{https://www.gem5.org/documentation/general_docs/building})
\end{itemize}

\subsection{Installation}

\begin{enumerate}
    \item Download and extract the Zenodo archive (DOI:~\texttt{10.5281/zenodo.18912932}).

    \item Build gem5:
    \begin{lstlisting}[
    language=Bash,
    basicstyle=\ttfamily\footnotesize,
    breaklines=true,
    breakatwhitespace=true
    ]
    cd gem5 && scons build/ALL/gem5.opt \ -j$(nproc)
    \end{lstlisting}

    \item Verify KVM is available on the host machine:
    \begin{lstlisting}[
    language=Bash,
    basicstyle=\ttfamily\footnotesize,
    breaklines=true,
    breakatwhitespace=true
    ]
    ls /dev/kvm
    \end{lstlisting}

\end{enumerate}

\subsection{Experiment workflow}

\begin{enumerate}

    \item Build the NPB disk image using the provided Ubuntu 24.04 base disk-image:
    \begin{lstlisting}[
    language=Bash,
    basicstyle=\ttfamily\footnotesize,
    breaklines=true,
    breakatwhitespace=true
    ]
    cd npb-24.04-imgs && \
    ./build-x86.sh
    \end{lstlisting}
    This script builds the NPB disk image from the locally provided base image, and automatically populates the \texttt{iso\_checksum} and \texttt{iso\_urls} fields in the Packer script.
    By default, it expects the base image at the relative path \path{../base-resources/x86-ubuntu-24.04-20250515}, which is the location of the base image that is provided in the Zenodo archive.
    A custom path can be provided as an argument if needed:
    \begin{lstlisting}[
    language=Bash,
    basicstyle=\ttfamily\footnotesize,
    breaklines=true,
    breakatwhitespace=true
    ]
    ./build-x86.sh /path/to/base-image
    \end{lstlisting}
    
    \item After the build completes, the NPB disk image will be available in the \texttt{npb-24.04-imgs/disk-image-x86-npb} directory.

    \item Run the \texttt{scripts/create\_resources.py} script to generate a \texttt{resources.json} file containing the correct paths and checksums for the NPB disk image, kernel, and a workload to run the NPB BT Class S simulation.
    By default, the script looks for the NPB disk-image in the \path{npb-24.04-imgs/disk-image-x86-npb} directory and the kernel in the \path{base-resources} directory.
    Custom paths can be provided as arguments if needed:
    \begin{lstlisting}[
    language=Bash,
    basicstyle=\ttfamily\footnotesize,
    breaklines=true,
    breakatwhitespace=true
    ]
    python3 scripts/create_resources.py \
        --disk-image /path/to/npb-disk-image \
        --kernel /path/to/kernel \
    \end{lstlisting}
    \item Run the simulation, with the following command. This command will use the \texttt{resources.json} file generated in the previous step as a local resources source in gem5:
\begin{lstlisting}[
language=Bash,
basicstyle=\ttfamily\scriptsize,
breaklines=true,
breakatwhitespace=true
]
GEM5_RESOURCE_JSON_APPEND=<path/to/resources.json> \
    ./gem5/build/ALL/gem5.opt \
    -d <out-dir> -re \
    ./disk-image-validator/disk-image-validate.py \
    --workload="x86-ubuntu-24.04-npb-bt-s-ispass-ae" \
    --resource_version="1.0.0" \
    --isa="x86"
\end{lstlisting}
Note: The \texttt{GEM5\_RESOURCE\_JSON\_APPEND} environment variable allows gem5 to use the resources defined in the provided \texttt{resources.json} file with \texttt{obtain\_resource} calls in the simulation script.
\\

    The script uses gem5's prebuilt x86 demo board. Stats are
    dumped at every hypercall event, the output directory
    will contain a \texttt{stats.txt} with
    exactly 5 stat dumps corresponding to the 5 hypercalls
    called by the workload.
\end{enumerate}

\subsection{Evaluation and expected results}

\paragraph{Hypercall execution order (Figure~\ref{fig:hypercall-execution-validation}).}
The simulation output (simout.txt) should show hypercalls triggered in the following
order: kernel boot~(1) $\rightarrow$ login~(2) $\rightarrow$ ROI
begin~(4) $\rightarrow$ ROI end~(5) $\rightarrow$ exit~(3),
matching Figure~\ref{fig:hypercall-execution-validation} of the paper.

\paragraph{ROI instruction count (Table~\ref{tab:benchmark-stats}).} 
Extract the ROI stats using the provided script:
\begin{lstlisting} [
language=Bash,
basicstyle=\ttfamily\footnotesize,
breaklines=true,
breakatwhitespace=true
]
python3 scripts/parse_stats.py --stats-file \ <path-to-stats.txt>
\end{lstlisting}
The extracted \texttt{ROIsimInsts} value should be approximately 495 million, matching the x86 NPB BT Size~S instruction count reported in Table~\ref{tab:benchmark-stats} of the paper.

\subsection{Experiment customization}
All other disk images (ARM, RISC-V, GAPBS) follow the same
Packer-based workflow described here:
\begin{itemize}
  \item For ARM: run \texttt{build-arm.sh} instead of
        \texttt{build-x86.sh}. ARM builds require an ARM host with
        KVM support. The \texttt{flash0.img} file required by the
        ARM Packer script is generated automatically by
        \texttt{build-arm.sh}.
  \item For RISC-V: run \texttt{build-riscv.sh}. KVM is not
        used but builds will be slower.
  \item For GAPBS: the build workflow is identical to NPB disk image,
        see the \texttt{gapbs/} directory in the Zenodo
        archive.
\end{itemize}

\subsection{Notes}

\begin{itemize}
  \item The NPB disk image build depends on the base image.
  \item Packer login timing may need adjustment on slow machines.
        See the note about \texttt{<wait>} commands in
        \texttt{npb-24.04-imgs/README.md}.
  \item MD5 checksums for locally built disk images can be computed
        with \texttt{md5sum <image>}. The exact checksum will differ
        from the pre-built images on gem5 Resources since the Packer
        build process is not bit-for-bit reproducible, but the
        simulation behavior should be identical.
\end{itemize}

\end{document}